\documentclass{book}
\usepackage{graphicx}
\usepackage{amsmath}
\usepackage{amsfonts}
\usepackage{amssymb}

\begin{document}

\frontmatter
\title{Introduction to Bosonic String Theory}
\author{Eduard Alexis Larra\~{n}aga Rubio}
\date{Bogot\'{a}, D.C. \ November 2002}
\maketitle
\tableofcontents

\chapter{\bigskip Introduction}

The development of fundamental physics in the past century arose from the
identification and the overcoming of contradictions between the existing
ideas. Fro example, the incompatibility of Maxwell's equations and Galilean
invariance led Einstein to create the Special Theory of Relativity, and the
inconsistency of this ones with Newtonian gravity produced the General Theory
of Relativity. Same thing happend with the conciliation between special
relativity and quantum mechanics, wich led to the development of Quantum Field Theory.

Now, there is another incompatibility: General Relativity and Quantum Field
theory. The quantization of gravity seems to be a non-renormalizable theory.

\bigskip

In the past year, String Theory has been the leading candidate for a theory
that unifies all fundamental forces in nature in a consistent scheme. The way
that string theory does this is to give up one of the basic assumptions of
quantum field theory, that elementary particles are mathematical points, and
instead to develop aquantum field theory of one-dimensional extended objects,
called \textit{strings}.

\bigskip

Ths string theory is still in progress, and there is not yet a complete
description of the standard model of elementary particles based on it.
However, there are some important features that seems to be gnereic in every
kind of string theory: The first, and most important, is that general
relativity is already incorporated in the theory. Although ordinary quantum
field theory does not allow gravity, string theory requires it. The second
fact is that the Yang-Mills gauge theories of the sort that comprise the
standard model naturally arise in string theory, but there is not yet a fully
understanding of why should we prefer the specific $SU\left(  3\right)
\otimes SU\left(  2\right)  \otimes U\left(  1\right)  $ gauge theory.

\bigskip

Here is given a little introduction to the most basic string theory: th
bosonic string. First there is given \ the basics of the classical string
theory as a generalization of the concept of world-line of a particle. Then
the bosonic string is constructed and the oscillation modes are described. The
covariant canonical quantization procedure is used and the bosonic spectra is shown.

\mainmatter

\chapter{\bigskip Classical String Theory}

\section{ World-Line of a Point Particle}

\bigskip

Classically, a point particle follow a trajectory in space time that is called
'world-line'. The world line can be expressed as functions $x^{\mu}\left(
\tau\right)  $\ with a parameter $\tau.$ The path is an extremal of the
action, that for a point particle \ of mass $m$ is proportional to the length
of the world line:%

\begin{equation}
S=m\int ds=m\int\sqrt{-\eta_{\mu\nu}\overset{.}{x}^{\mu}\overset{.}{x}^{v}%
}d\tau\label{point particle action}%
\end{equation}

where $\eta_{\mu\nu}$ is the flat Minkowski space-time metric. The momentum
conjugate to $x^{\mu}\left(  \tau\right)  $ is:%

\begin{equation}
p_{\mu}=-\frac{\partial L}{\partial\overset{.}{x}^{\mu}}=-\frac{\partial
\left(  m\sqrt{-\overset{.}{x}^{\mu}\overset{.}{x}_{\mu}}\right)  }%
{\partial\overset{.}{x}^{\mu}}=m\frac{\overset{.}{x}_{\mu}}{\sqrt{-\overset
{.}{x}^{\mu}\overset{.}{x}_{\mu}}}%
\end{equation}%

\begin{equation}
p_{\mu}=m\frac{\overset{.}{x}_{\mu}}{\sqrt{-\overset{.}{x}^{2}}}%
\end{equation}

\bigskip

This equation gives the constrain of mass-shell:%

\begin{equation}
p^{2}=-m^{2}\label{constrain}%
\end{equation}

The Lagrange equation from varying $S$ with respect to $x\left(  \tau\right)
$ is:%

\begin{equation}
\frac{\partial L}{\partial x^{\mu}}-\frac{\partial}{\partial\tau}\left(
\frac{\partial L}{\partial\overset{.}{x}^{\mu}}\right)  =0
\end{equation}%

\begin{equation}
\frac{\partial}{\partial\tau}\left(  \frac{m\overset{.}{x}_{\mu}}%
{\sqrt{-\overset{.}{x}^{2}}}\right)  =0\label{equation point particle}%
\end{equation}

\bigskip

The canonical Hamiltonian is:%

\begin{equation}
H_{can}=\frac{\partial L}{\partial\overset{.}{x}^{\mu}}\overset{.}{x}^{\mu
}-L=\frac{m\overset{.}{x}_{\mu}\overset{.}{x}^{\mu}}{\sqrt{-\overset{.}%
{x}^{\mu}\overset{.}{x}_{\mu}}}-m\sqrt{-\overset{.}{x}^{\mu}\overset{.}%
{x}_{\mu}}%
\end{equation}%

\begin{equation}
H_{can}=0
\end{equation}

So, $H_{can}$ vanishes identically. This means that the constrain given in
(\ref{constrain}) governs the dynamics of the system completly. This constrain
can be added to the Hamiltonian by using a Lagrange multiplier. This gives:%

\begin{equation}
H=\frac{N}{2m}\left(  p^{2}+m^{2}\right)
\end{equation}

\bigskip

The equation of motion in Poisson bracket notation is:%

\begin{equation}
\overset{.}{x}^{\mu}=\left\{  x^{\mu},H\right\}  =\frac{\partial H}{\partial
p_{\mu}}=\frac{N}{2m}\frac{\partial\left(  p^{2}+m^{2}\right)  }{\partial
p_{\mu}}%
\end{equation}%

\begin{equation}
\overset{.}{x}^{\mu}=\frac{N}{m}p^{\mu}=N\frac{\overset{.}{x}^{\mu}}%
{\sqrt{-\overset{.}{x}^{2}}}%
\end{equation}

\bigskip

So, it gives:%

\begin{equation}
\overset{.}{x}^{2}=-N^{2}%
\end{equation}

\bigskip i.e. we are describing time-like trajectories.

The action in (\ref{point particle action}) is invariant under local
reparametrizations. This kind of gauge invariance is represented by a change
$\tau\longrightarrow\tau\left(  \widetilde{\tau}\right)  $. This
reparametrization invariance is a one-dimensional analog of the four
dimensional general coordinate invariance. The choice $N=1$ corresponds to the
called ''static gauge'':%

\begin{equation}
x^{0}=\tau
\end{equation}

And in this gauge the action becomes:%

\begin{equation}
S=m\int\sqrt{1-v^{2}}dt
\end{equation}

\bigskip

where the parameter is renamed $t$ and:%

\begin{equation}
\overrightarrow{v}=\frac{d\overrightarrow{x}}{dt}%
\end{equation}
\bigskip

The equation of motion becomes:%

\begin{equation}
\frac{d\overrightarrow{p}}{dt}=0
\end{equation}

with:%

\begin{equation}
\overrightarrow{p}=\frac{m\overrightarrow{v}}{\sqrt{1-v^{2}}}%
\end{equation}

\subsection{\bigskip Alternative Action for a point particle}

An alternative action for the point particle is:%

\begin{equation}
S=-\frac{1}{2}\int\left[  \eta_{\mu\nu}\overset{.}{x}^{\mu}\overset{.}{x}%
^{v}e^{-1}\left(  \tau\right)  -m^{2}e\left(  \tau\right)  \right]
d\tau\label{einbein action}%
\end{equation}

\bigskip

where $e\left(  \tau\right)  $ (called the 'einbein') is a new independient
function. This action allows to make the generalization to the massless case,
and doesn't have square roots like in (\ref{point particle action}) that makes
the treatment of the quantum theory complicated. In order to prove that the
action are equivalent, lets \ make the variation of $e\left(  \tau\right)  $:%

\begin{equation}
\delta S=\frac{1}{2}\int\left[  \eta_{\mu\nu}\overset{.}{x}^{\mu}\overset
{.}{x}^{v}e^{-2}\left(  \tau\right)  +m^{2}\right]  \delta e\left(
\tau\right)  d\tau
\end{equation}

\bigskip

Then, setting $\delta S=0$ we obtain:%

\begin{equation}
\eta_{\mu\nu}\overset{.}{x}^{\mu}\overset{.}{x}^{v}e^{-2}\left(  \tau\right)
+m^{2}=0
\end{equation}%

\begin{equation}
e\left(  \tau\right)  =\frac{1}{m}\sqrt{-\eta_{\mu\nu}\overset{.}{x}^{\mu
}\overset{.}{x}^{v}}\label{einbein}%
\end{equation}

\bigskip

If we substitute (\ref{einbein}) in (\ref{einbein action}) we get:%

\begin{equation}
S=-\frac{1}{2}\int\left[  \eta_{\mu\nu}\overset{.}{x}^{\mu}\overset{.}{x}%
^{v}\left(  \frac{1}{m}\sqrt{-\eta_{\mu\nu}\overset{.}{x}^{\mu}\overset{.}%
{x}^{v}}\right)  ^{-1}-m^{2}\left(  \frac{1}{m}\sqrt{-\eta_{\mu\nu}\overset
{.}{x}^{\mu}\overset{.}{x}^{v}}\right)  \right]  d\tau
\end{equation}%

\begin{equation}
S=-\frac{1}{2}\int\left[  -m\sqrt{-\eta_{\mu\nu}\overset{.}{x}^{\mu}%
\overset{.}{x}^{v}}-m\sqrt{-\eta_{\mu\nu}\overset{.}{x}^{\mu}\overset{.}%
{x}^{v}}\right]  d\tau
\end{equation}%

\begin{equation}
S=\int m\sqrt{-\eta_{\mu\nu}\overset{.}{x}^{\mu}\overset{.}{x}^{v}}d\tau
\end{equation}

\bigskip Hence, the actions are equivalent. Varying $x^{\mu}$ in (\ref{einbein
action}) we get:%

\begin{equation}
\delta S=-\frac{1}{2}\int\left[  2\overset{.}{x}^{\mu}e^{-1}\left(
\tau\right)  \right]  \delta\left(  \overset{.}{x}^{\mu}\right)  d\tau
\end{equation}%

\begin{equation}
\delta S=-\int\left[  \overset{.}{x}^{\mu}e^{-1}\left(  \tau\right)  \right]
\partial_{\tau}\delta x^{\mu}d\tau
\end{equation}

Integrating by parts we obtain:%

\begin{equation}
\delta S=-\left[  \overset{.}{x}^{\mu}e^{-1}\left(  \tau\right)  \right]
\delta x^{\mu}+\int\partial_{\tau}\left[  \overset{.}{x}^{\mu}e^{-1}\left(
\tau\right)  \right]  \delta x^{\mu}d\tau
\end{equation}

\bigskip

since at the extremes the variation is zero, then:%

\begin{equation}
\delta S=\int\partial_{\tau}\left[  \overset{.}{x}^{\mu}e^{-1}\left(
\tau\right)  \right]  \delta x^{\mu}d\tau
\end{equation}

\bigskip

Then, setting $\delta S=0$ gives:%

\begin{equation}
\partial_{\tau}\left[  \overset{.}{x}^{\mu}e^{-1}\left(  \tau\right)  \right]
=0
\end{equation}

\bigskip Substituing (\ref{einbein}) we get:%

\begin{equation}
\partial_{\tau}\left[  \overset{.}{x}^{\mu}\left(  \frac{1}{m}\sqrt{-\eta
_{\mu\nu}\overset{.}{x}^{\mu}\overset{.}{x}^{v}}\right)  ^{-1}\right]  =0
\end{equation}%

\begin{equation}
\partial_{\tau}\left[  \frac{m\overset{.}{x}^{\mu}}{\sqrt{-\eta_{\mu\nu
}\overset{.}{x}^{\mu}\overset{.}{x}^{v}}}\right]  =0
\end{equation}

\bigskip

and this is the same equation of motion as (\ref{equation point particle}).

\bigskip

\section{World-Volume Action}

\bigskip

As we have seen, the action for a point particle was proportional to the
lenght of its world-line. For strings, the action will be prportional to the
surface area of its world-sheet, and in general for a $p-$brane the action
involves the $\left(  p+1\right)  -$dimensional volume:%

\begin{equation}
S=-T_{p}\int d\mu_{p+1}\label{world-volume action}%
\end{equation}

where the\ constant $T_{p}$ makes the action dimensionless, so it has
dimensions of \ [mass]$^{\left(  p+1\right)  }$ or [lenght]$^{-\left(
p+1\right)  }$ . This factor is asociated with the tension of the $p-$brane.
For a $0-$ brane, (i.e. a point particle) it is just the mass.

\bigskip

Suppose that $\xi^{\alpha}$ $\ \left(  \alpha=0,1,2,...,p+1\right)  $ are the
coordinates in the world-volume of the $p-$brane, and $g_{\mu\nu}$ $\left(
\mu,\nu=0,1,2,....,d-1\right)  $ is the metric of the $d-$dimensional
space-time in which the $p-$brane propagates. Then, $g_{\mu\nu} $ induces a
metric in the world-volume given by:%

\begin{equation}
ds^{2}=-g_{\mu\nu}dx^{\mu}dx^{\nu}=-g_{\mu\nu}\frac{\partial x^{\mu}}%
{\partial\xi^{\alpha}}\frac{\partial x^{\nu}}{\partial\xi^{\beta}}d\xi
^{\alpha}d\xi^{\beta}=G_{\alpha\beta}d\xi^{\alpha}d\xi^{\beta}%
\end{equation}

\bigskip where the induced metric is:
\begin{equation}
G_{\alpha\beta}=-g_{\mu\nu}\frac{\partial x^{\mu}}{\partial\xi^{\alpha}}%
\frac{\partial x^{\nu}}{\partial\xi^{\beta}}%
\end{equation}

Then, the invariant volume element is given by:%

\begin{equation}
d\mu_{p+1}=\sqrt{-\det G_{\alpha\beta}}d^{p+1}\xi
\end{equation}

\section{\bigskip Nambu-Goto Action}

In the case of $1-$branes, i.e. strings, the world-volume action
(\ref{world-volume action}) becomes:%

\begin{equation}
S=-T\int\sqrt{-\det G_{\alpha\beta}}d^{2}\xi
\end{equation}

\bigskip

If the space-time is flat Minkowski $g_{\mu\nu}=\eta_{\mu\nu}$ we have:%

\begin{equation}
\sqrt{-\det G_{\alpha\beta}}=\sqrt{-\det\left(  -\eta_{\mu\nu}\frac{\partial
x^{\mu}}{\partial\xi^{\alpha}}\frac{\partial x^{\nu}}{\partial\xi^{\beta}%
}\right)  }%
\end{equation}

Defining $\xi^{0}=\tau$ and $\xi^{1}=\sigma,$ then:%

\begin{equation}
\sqrt{-\det G_{\alpha\beta}}=\sqrt{\left(  \overset{.}{x}\cdot x^{\prime
}\right)  ^{2}-\overset{.}{x}^{2}x^{\prime2}}%
\end{equation}

where:%

\begin{equation}
\overset{.}{x}=\frac{\partial x}{\partial\tau}\text{ \ \ \ \ \ \ \ \ \ \ }%
x^{\prime}=\frac{\partial x}{\partial\sigma}\text{\ \ }%
\end{equation}
\ 

\bigskip

and the action for the string, called Nambu-Goto action becomes:%

\begin{equation}
S_{NG}=-T\int\sqrt{\left(  \overset{.}{x}\cdot x^{\prime}\right)
^{2}-\overset{.}{x}^{2}x^{\prime2}}d\sigma d\tau\label{nambu goto action}%
\end{equation}

\bigskip

\section{Polyakov Action}

\bigskip

Again, as in the point particle case, the square root in the Nambu-Goto action
make the quantum treatment complicated. So, we introduce an equivalent action:%

\begin{equation}
S_{P}=-\frac{T}{2}\int\sqrt{-\det h_{\alpha\beta}}h^{\alpha\beta}\eta_{\mu\nu
}\partial_{\alpha}x^{\mu}\partial_{\beta}x^{\nu}d^{2}\xi
\label{polyakov action}%
\end{equation}

\bigskip

where $h_{\alpha\beta}\left(  \sigma,\tau\right)  $ is the world-sheet metric
and $\partial_{\alpha}x^{\mu}=\frac{\partial x^{\mu}}{\partial\xi^{\alpha}}$.
The stress-tensor is defined \ as the variation of the action with respect to
the metric:%

\begin{equation}
T_{\alpha\beta}\equiv-\frac{2}{T}\frac{1}{\sqrt{-\det h_{\alpha\beta}}}%
\frac{\delta S}{\delta h^{\alpha\beta}}%
\end{equation}%

\begin{equation}
T_{\alpha\beta}=\eta_{\mu\nu}\partial_{\alpha}x^{\mu}\partial_{\beta}x^{\nu
}-\frac{1}{2}h_{\alpha\beta}h^{\gamma\delta}\eta_{\mu\nu}\partial_{\gamma
}x^{\mu}\partial_{\delta}x^{\nu}%
\end{equation}

\bigskip Hence, the Euler-Lagrange equation is:%

\begin{equation}
\partial_{\alpha}x\cdot\partial_{\beta}x-\frac{1}{2}h_{\alpha\beta}%
h^{\gamma\delta}\partial_{\gamma}x\cdot\partial_{\delta}%
x=0\label{eulerlagrange polyakov}%
\end{equation}

\bigskip

Solving for $h_{\alpha\beta}$ we obtain:%

\begin{equation}
h_{\alpha\beta}=\partial_{\alpha}x\cdot\partial_{\beta}x=\eta_{\mu\nu}%
\partial_{\alpha}x^{\mu}\partial_{\beta}x^{\nu}\label{metric equvalence}%
\end{equation}

\bigskip

Substituing this in (\ref{eulerlagrange polyakov}) gives:%

\begin{equation}
h_{\alpha\beta}-\frac{1}{2}h_{\alpha\beta}h^{\gamma\delta}h_{\gamma\delta
}=h_{\alpha\beta}-\frac{1}{2}h_{\alpha\beta}\delta_{\gamma}^{\gamma}%
=h_{\alpha\beta}-h_{\alpha\beta}=0
\end{equation}

\bigskip

So, (\ref{metric equvalence}) means that the metric in the world-sheet
$h_{\alpha\beta}$ is equal (at least classically) to the induced metric.
Substituing this in the Polyakov action we obtain:%

\begin{equation}
S_{P}=-\frac{T}{2}\int\sqrt{-\det\left(  \eta_{\mu\nu}\partial_{\alpha}x^{\mu
}\partial_{\beta}x^{\nu}\right)  }h^{\alpha\beta}h_{\alpha\beta}d^{2}\xi
\end{equation}%

\begin{equation}
S_{P}=-\frac{T}{2}\int\sqrt{-\det\left(  \eta_{\mu\nu}\partial_{\alpha}x^{\mu
}\partial_{\beta}x^{\nu}\right)  }\delta_{\alpha}^{\alpha}d^{2}\xi
\end{equation}%

\begin{equation}
S_{P}=-T\int\sqrt{-\det\left(  \eta_{\mu\nu}\partial_{\alpha}x^{\mu}%
\partial_{\beta}x^{\nu}\right)  }d^{2}\xi=S_{NG}%
\end{equation}

\bigskip

Hence the Polyakov and the Nambu-Goto actions are equivalent, at least
classically, since in the quantum treatment it is not in general.

\bigskip

\section{Symmetries of the Polyakov Action}

\bigskip

Polyakov action has the symmetries:

\begin{enumerate}
\item \bigskip Poincar\'{e} invariance:
\begin{equation}
x^{\mu}\longmapsto\omega_{\text{ \ }\nu}^{\mu}x^{\nu}+a^{\mu}\text{ \ \ \ \ ,
\ \ \ \ }h_{\alpha\beta}\longmapsto h_{\alpha\beta}%
\end{equation}

with $\omega_{\mu\nu}=-\omega_{\nu\mu}$.

\item  Local \ 2-dimensional reparametrization invariance%

\begin{equation}
\xi^{\alpha}\longmapsto\xi^{\prime\alpha}\left(  \xi^{\beta}\right)
\end{equation}

\item  Conformal or Weyl invariance%

\begin{equation}
h_{\alpha\beta}\longmapsto\Lambda\left(  \xi^{\gamma}\right)  h_{\alpha\beta
}\text{ \ \ \ \ , \ \ \ \ }x^{\mu}=x^{\mu}%
\end{equation}
\end{enumerate}

The two reparametrization invariance symmetries of $S$ allow us to choose a
gauge in which the three independient components of $h_{\alpha\beta}$ are
expressed with just one function. The usual choose is th conformal flat gauge:%

\begin{equation}
h_{\alpha\beta}=e^{\Lambda\left(  \xi\right)  }\eta_{\alpha\beta}%
\end{equation}

\bigskip

where $\eta_{\alpha\beta}=\left(
\begin{array}
[c]{cc}%
-1 & 0\\
0 & 1
\end{array}
\right)  $ is the 2-dimensional Minkowski metric for a flat world-sheet.
Substituing this in the action que obtain:%

\begin{equation}
S=-\frac{T}{2}\int\sqrt{-\det\left(  e^{\Lambda\left(  \xi\right)  }%
\eta_{\alpha\beta}\right)  }e^{-\Lambda\left(  \xi\right)  }\eta^{\alpha\beta
}\eta_{\mu\nu}\partial_{\alpha}x^{\mu}\partial_{\beta}x^{\nu}d^{2}\xi
\end{equation}%

\begin{equation}
S=-\frac{T}{2}\int\sqrt{e^{2\Lambda\left(  \xi\right)  }}e^{-\Lambda\left(
\xi\right)  }\eta^{\alpha\beta}\eta_{\mu\nu}\partial_{\alpha}x^{\mu}%
\partial_{\beta}x^{\nu}d^{2}\xi
\end{equation}%

\begin{equation}
S=-\frac{T}{2}\int\eta^{\alpha\beta}\eta_{\mu\nu}\partial_{\alpha}x^{\mu
}\partial_{\beta}x^{\nu}d^{2}\xi\label{gauge fixed polyakov action}%
\end{equation}

As can be seen, the gauge fixed action is quadratic in the $x$'s. Thus,
mathematically \ it is the same as a theory of $d$ free scalar fields in two
dimensions. Varying $x^{\mu}$ we obtain the equation of motion:%

\begin{equation}
\overset{..}{x}^{\mu}-x^{\prime\prime\mu}=0\label{wave equation}%
\end{equation}

\bigskip

This is simply a free two dimensional wave equation. Also, we have to take
account of the constraints $T_{\alpha\beta}=0$. In the conformal gauge this
constrain is:%

\begin{equation}
T_{\alpha\beta}=\eta_{\mu\nu}\partial_{\alpha}x^{\mu}\partial_{\beta}x^{\nu
}-\frac{1}{2}\eta_{\alpha\beta}\eta^{\gamma\delta}\eta_{\mu\nu}\partial
_{\gamma}x^{\mu}\partial_{\delta}x^{\nu}=0
\end{equation}

This can be writen as:%

\begin{equation}
T_{00}=\eta_{\mu\nu}\overset{.}{x}^{\mu}\overset{.}{x}^{\nu}+\frac{1}{2}%
\eta^{\gamma\delta}\eta_{\mu\nu}\partial_{\gamma}x^{\mu}\partial_{\delta
}x^{\nu}=0
\end{equation}%

\begin{equation}
T_{00}=\eta_{\mu\nu}\overset{.}{x}^{\mu}\overset{.}{x}^{\nu}-\frac{1}{2}%
\eta_{\mu\nu}\overset{.}{x}^{\mu}\overset{.}{x}^{\nu}+\frac{1}{2}\eta_{\mu\nu
}x^{\prime\mu}x^{\prime\nu}=0
\end{equation}%

\begin{equation}
T_{00}=T_{11}=\frac{1}{2}\left(  \eta_{\mu\nu}\overset{.}{x}^{\mu}\overset
{.}{x}^{\nu}+\eta_{\mu\nu}x^{\prime\mu}x^{\prime\nu}\right)
=0\label{constraint1}%
\end{equation}%

\begin{equation}
T_{01}=T_{10}=\eta_{\mu\nu}\overset{.}{x}^{\mu}x^{\prime\nu}%
=0\label{constraint2}%
\end{equation}

\bigskip

Adding and substracting (\ref{constraint1}) and (\ref{constraint2}) gives the condition:%

\begin{equation}
\left(  \overset{.}{x}\pm x^{\prime}\right)  ^{2}=0\label{visaroro constraint}%
\end{equation}

\bigskip Known as the Visaroro constraints, and they are the analog of the
Gauss law in the string case.

\section{\bigskip Light-cone coordinates}

Define the light-cone coordinates as:%

\begin{equation}
\xi_{\pm}=\tau\pm\sigma
\end{equation}

\bigskip

The flat metric becomes:%

\begin{equation}
ds^{2}=-d\tau^{2}+d\sigma^{2}=-\left(  d\tau+d\sigma\right)  \left(
d\tau-d\sigma\right)  =-d\xi_{+}d\xi_{-}%
\end{equation}

\bigskip

Thus, the metric components are:%

\begin{equation}
g_{++}=g_{--}=0\text{ \ \ \ \ , \ \ \ \ }g_{+-}=g_{-+}=-\frac{1}{2}%
\end{equation}

and \ we have:%

\begin{equation}
\partial_{\pm}=\frac{1}{2}\left(  \partial_{\tau}\pm\partial_{\sigma}\right)
\end{equation}

\bigskip Then, the Polyakov action in the conformal gauge (\ref{gauge fixed
polyakov action}) may be writen as:%

\begin{equation}
S=2T\int\eta_{\mu\nu}\partial_{+}x^{\mu}\partial_{-}x^{\nu}d^{2}%
\xi\label{polyakov action light coordinates}%
\end{equation}

\bigskip

The equations of motion from the Polyakov action (\ref{wave equation}) becomes:%

\begin{equation}
\overset{..}{x}^{\mu}-x^{\prime\prime\mu}=\partial_{\tau}\partial_{\tau}%
x^{\mu}-\partial_{\sigma}\partial_{\sigma}x^{\mu}=0
\end{equation}%

\begin{equation}
\left(  \partial_{\tau}+\partial_{\sigma}\right)  \left(  \partial_{\tau
}x^{\mu}-\partial_{\sigma}x^{\mu}\right)  =0
\end{equation}%

\begin{equation}
\left(  \partial_{\tau}+\partial_{\sigma}\right)  \left(  \partial_{\tau
}-\partial_{\sigma}\right)  x^{\mu}=0
\end{equation}%

\begin{equation}
\partial_{+}\partial_{-}x^{\mu}=0\label{equation of motion light coordinates}%
\end{equation}

The stress-tensor in light-coordinates is:%

\begin{equation}
T_{++}=\frac{1}{2}\eta_{\mu\nu}\partial_{+}x^{\mu}\partial_{+}x^{\nu}=\frac
{1}{2}\eta_{\mu\nu}\left(  \overset{.}{x}^{\mu}+x^{\prime\mu}\right)  \left(
\overset{.}{x}^{\nu}+x^{\prime\nu}\right)  =T_{00}+2T_{10}%
\end{equation}%

\begin{equation}
T_{--}=\frac{1}{2}\eta_{\mu\nu}\partial_{-}x^{\mu}\partial_{-}x^{\nu}=\frac
{1}{2}\eta_{\mu\nu}\left(  \overset{.}{x}^{\mu}-x^{\prime\mu}\right)  \left(
\overset{.}{x}^{\nu}-x^{\prime\nu}\right)  =T_{00}-2T_{10}%
\end{equation}%

\begin{equation}
T_{+-}=T_{-+}=0\label{no1}%
\end{equation}

\bigskip In these coordinates, the Energy-momentum conservation is given by:%

\begin{equation}
\nabla^{\alpha}T_{\alpha\beta}=-\frac{1}{2}\left(  \partial_{-}T_{++}%
+\partial_{+}T_{-+}\right)  =-\frac{1}{2}\left(  \partial_{+}T_{--}%
+\partial_{-}T_{+-}\right)  =0
\end{equation}

but using (\ref{no1}) we obtain:%

\begin{equation}
\partial_{-}T_{++}=\partial_{+}T_{--}=0
\end{equation}

\bigskip And finally, the Visaroro constraints (\ref{visaroro constraint}) can
be written as:%

\begin{equation}
T_{--}=\frac{1}{2}\eta_{\mu\nu}\partial_{-}x^{\mu}\partial_{-}x^{\nu
}=0\text{\ \ \ \ \ \ \ , \ \ \ \ \ \ }T_{++}=\frac{1}{2}\eta_{\mu\nu}%
\partial_{+}x^{\mu}\partial_{+}x^{\nu}%
=0\label{visaroro constraint light coordinates}%
\end{equation}

\section{\bigskip Boundary Conditions}

\bigskip In string theory there are three important types:

In the case of \textit{closed strings} the world-sheet is a tube, and one
should impose periodicity in the spacial parameter $\sigma$. Usually the
period is choosen to be $\pi$:%

\begin{equation}
x^{\mu}\left(  \sigma+\pi,\tau\right)  =x^{\mu}\left(  \sigma,\tau\right)
\end{equation}

\bigskip For \textit{open strings} the world-sheet is a strip. The conditions
used are:%

\begin{equation}
\text{Neumann: \ \ \ }\frac{\partial x^{\mu}}{\partial\sigma}=0\text{
\ \ \ \ at \ }\sigma=0\text{ or }\pi
\end{equation}%

\begin{equation}
\text{Dirichlet: \ \ \ \ }\frac{\partial x^{\mu}}{\partial\tau}=0\text{
\ \ \ \ \ at \ }\sigma=0\text{ or }\pi
\end{equation}

Neumann condition imply that no momentum flows off the ends of the string.
Dirichlet condition implies that the end points of the string are fixed in
spacetime, and that they ends on a physical object called $D-$brane ( D stands
for Dirichlet).

\bigskip

\bigskip

\section{Oscillator expansions}

\bigskip

As we have seen the equation of motion for the string is (\ref{wave equation}):%

\begin{equation}
\overset{..}{x}^{\mu}-x^{\prime\prime\mu}=0
\end{equation}

but we must treat each case of boundary condition separately:

\bigskip

\begin{enumerate}
\item \textbf{Closed Strings}

For closed strings the general solution of the two-dimensional wave equation
is a sum of 'right' movers and 'left' movers:%

\begin{equation}
x^{\mu}\left(  \sigma,\tau\right)  =x_{R}^{\mu}\left(  \tau-\sigma\right)
+x_{L}^{\mu}\left(  \tau+\sigma\right) \label{solution closed strings}%
\end{equation}

This solution must satisfy the conditions:

\begin{itemize}
\item $x^{\mu}\left(  \sigma,\tau\right)  $ \ is real

\item $x^{\mu}\left(  \sigma+\pi,\tau\right)  =x^{\mu}\left(  \sigma
,\tau\right)  $
\end{itemize}

This condition can be solved in terms of Fourier series as:%

\begin{equation}
x_{R}^{\mu}\left(  \tau-\sigma\right)  =\frac{1}{2}x_{o}^{\mu}+\frac{1}{2\pi
T}p^{\mu}\left(  \tau-\sigma\right)  +\frac{i}{\sqrt{4\pi T}}\underset{n\neq
0}{\sum}\frac{1}{n}\alpha_{n}^{\mu}e^{-in\left(  \tau-\sigma\right)
}\label{right movers}%
\end{equation}%

\begin{equation}
x_{L}^{\mu}\left(  \tau+\sigma\right)  =\frac{1}{2}x_{o}^{\mu}+\frac{1}{2\pi
T}p^{\mu}\left(  \tau+\sigma\right)  +\frac{i}{\sqrt{4\pi T}}\underset{n\neq
0}{\sum}\frac{1}{n}\widetilde{\alpha}_{n}^{\mu}e^{-in\left(  \tau
+\sigma\right)  }\label{left movers}%
\end{equation}

where $\alpha_{n}^{\mu}$ and $\widetilde{\alpha}_{n}^{\mu}$ are Fourier modes
that must satisfy (in order to $x^{\mu}\left(  \sigma,\tau\right)  $ to be real):%

\begin{equation}
\left(  \alpha_{n}^{\mu}\right)  ^{\dagger}=\alpha_{-n}^{\mu}\text{
\ \ \ \ \ \ \ and \ \ \ \ \ \ \ }\left(  \widetilde{\alpha}_{n}^{\mu}\right)
^{\dagger}=\widetilde{\alpha}_{-n}^{\mu}\label{alfa's constraints}%
\end{equation}

The center of mass coordinate $x^{\mu}$ and the momentum $p^{\mu}$ are real
too, and the factor :%

\begin{equation}
\ell_{s}=\frac{1}{\sqrt{2\pi T}}%
\end{equation}

is called the fundamental string lenght scale. So, the solution is:%

\begin{equation}
x^{\mu}\left(  \sigma,\tau\right)  =x_{o}^{\mu}+\frac{1}{\pi T}p^{\mu}%
\tau+\frac{i}{\sqrt{4\pi T}}\underset{n\neq0}{\sum}\frac{1}{n}\left(
\alpha_{n}^{\mu}e^{in\sigma}+\widetilde{\alpha}_{n}^{\mu}e^{-in\sigma}\right)
e^{-in\tau}\label{closed string solution}%
\end{equation}

Now, defining%

\begin{equation}
\alpha_{0}^{\mu}=\widetilde{\alpha}_{0}^{\mu}=\frac{1}{\sqrt{\pi T}}p^{\mu}%
\end{equation}

we can write in light-cone coordinates:%

\begin{equation}
\partial_{-}x^{\mu}\left(  \sigma,\tau\right)  =\partial_{-}x_{R}^{\mu}%
=\frac{1}{\sqrt{4\pi T}}\underset{n}{\sum}\alpha_{n}^{\mu}e^{-in\left(
\tau-\sigma\right)  }%
\end{equation}%
\begin{equation}
\partial_{+}x^{\mu}\left(  \sigma,\tau\right)  =\partial_{+}x_{L}^{\mu}%
=\frac{1}{\sqrt{4\pi T}}\underset{n}{\sum}\widetilde{\alpha}_{n}^{\mu
}e^{-in\left(  \tau+\sigma\right)  }%
\end{equation}

\item \textbf{Open Strings}

In the case of open strings we will use the Neumann boundary conditions:%

\begin{equation}
\frac{\partial x^{\mu}}{\partial\sigma}=0\text{ \ \ \ \ at \ }\sigma=0\text{
or }\pi
\end{equation}

Using the solution (\ref{solution closed strings}) with the expresion for
right and left movers (\ref{right movers}) and (\ref{left movers}) and
substituing in the \ Neumann condition we get:%

\begin{equation}
\left.  \frac{\partial x^{\mu}}{\partial\sigma}\right|  _{\sigma=0}=\left.
\frac{\partial x_{R}^{\mu}}{\partial\sigma}\right|  _{\sigma=0}+\left.
\frac{\partial x_{L}^{\mu}}{\partial\sigma}\right|  _{\sigma=0}=0
\end{equation}%
\begin{equation}
-\frac{1}{2\pi T}p^{\mu}-\frac{1}{\sqrt{4\pi T}}\underset{n\neq0}{\sum}%
\alpha_{n}^{\mu}e^{-in\tau}+\frac{1}{2\pi T}\widetilde{p}^{\mu}+\frac{1}%
{\sqrt{4\pi T}}\underset{n\neq0}{\sum}\widetilde{\alpha}_{n}^{\mu}e^{-in\tau
}=0
\end{equation}%
\begin{equation}
\frac{1}{2\pi T}\left(  \widetilde{p}^{\mu}-p^{\mu}\right)  +\frac{1}%
{\sqrt{4\pi T}}\underset{n\neq0}{\sum}\left(  \widetilde{\alpha}_{n}^{\mu
}-\alpha_{n}^{\mu}\right)  e^{-in\tau}=0
\end{equation}

Then we obtain:%

\begin{equation}
\widetilde{p}^{\mu}=p^{\mu}%
\end{equation}%
\begin{equation}
\widetilde{\alpha}_{n}^{\mu}=\alpha_{n}^{\mu}%
\end{equation}

This means that the right and left movers get mixed in the solution for the
open string. Now, using the boundary condition at $\sigma=\pi$ we get:%

\begin{equation}
\left.  \frac{\partial x^{\mu}}{\partial\sigma}\right|  _{\sigma=\pi}=\left.
\frac{\partial x_{R}^{\mu}}{\partial\sigma}\right|  _{\sigma=\pi}+\left.
\frac{\partial x_{L}^{\mu}}{\partial\sigma}\right|  _{\sigma=\pi}=0
\end{equation}%
\begin{equation}
-\frac{1}{2\pi T}p^{\mu}\pi-\frac{1}{\sqrt{4\pi T}}\underset{n\neq0}{\sum
}\alpha_{n}^{\mu}e^{-in\left(  \tau-\pi\right)  }+\frac{1}{2\pi T}p^{\mu}%
\pi+\frac{1}{\sqrt{4\pi T}}\underset{n\neq0}{\sum}\alpha_{n}^{\mu
}e^{-in\left(  \tau+\pi\right)  }=0
\end{equation}%
\begin{equation}
-\frac{1}{\sqrt{4\pi T}}\underset{n\neq0}{\sum}\alpha_{n}^{\mu}e^{-in\left(
\tau-\pi\right)  }+\frac{1}{\sqrt{4\pi T}}\underset{n\neq0}{\sum}\alpha
_{n}^{\mu}e^{-in\left(  \tau+\pi\right)  }=0
\end{equation}%
\begin{equation}
\frac{1}{\sqrt{4\pi T}}\underset{n\neq0}{\sum}\alpha_{n}^{\mu}\left(
e^{-in\pi}-e^{in\pi}\right)  =0
\end{equation}

Then, we get the condition:
\begin{equation}
e^{-in\pi}-e^{in\pi}=0
\end{equation}%
\begin{equation}
-2i\sin\left(  n\pi\right)  =0
\end{equation}

So, $n$ must be an integer. The solution now becomes:%

\begin{equation}
x^{\mu}\left(  \tau,\sigma\right)  =x_{o}^{\mu}+\frac{p^{\mu}}{\pi T}%
\tau+\frac{i}{\sqrt{4\pi T}}\underset{n\neq0}{\sum}\frac{1}{n}\alpha_{n}^{\mu
}e^{-in\tau}\left(  e^{in\sigma}+e^{-in\sigma}\right)
\end{equation}%
\begin{equation}
x^{\mu}\left(  \tau,\sigma\right)  =x_{o}^{\mu}+\frac{p^{\mu}}{\pi T}%
\tau+\frac{i}{\sqrt{4\pi T}}\underset{n\neq0}{\sum}\frac{1}{n}\alpha_{n}^{\mu
}e^{-in\tau}2\cos\left(  n\sigma\right)
\end{equation}%
\begin{equation}
x^{\mu}\left(  \tau,\sigma\right)  =x_{o}^{\mu}+\frac{p^{\mu}}{\pi T}%
\tau+\frac{i}{\sqrt{\pi T}}\underset{n\neq0}{\sum}\frac{1}{n}\alpha_{n}^{\mu
}e^{-in\tau}\cos\left(  n\sigma\right)
\end{equation}

Defining again $\alpha_{0}^{\mu}=\frac{1}{\sqrt{\pi T}}p^{\mu}$ we get:%

\begin{equation}
\partial_{\pm}x^{\mu}\left(  \tau,\sigma\right)  =\frac{1}{2\pi T}p^{\mu
}+\frac{1}{\sqrt{4\pi T}}\underset{n\neq0}{\sum}\alpha_{n}^{\mu}e^{-in\left(
\tau\pm\sigma\right)  }%
\end{equation}%
\begin{equation}
\partial_{\pm}x^{\mu}\left(  \tau,\sigma\right)  =\frac{1}{\sqrt{4\pi T}%
}\underset{n}{\sum}\alpha_{n}^{\mu}e^{-in\left(  \tau\pm\sigma\right)  }%
\end{equation}
\end{enumerate}

\bigskip

\section{Center of mass Position and Momentum}

The position and momentum of the center of mass for the open and closed
strings can be calculated, and they are given by:%

\begin{equation}
X_{CM}^{\mu}=\frac{1}{\pi}\int_{0}^{\pi}x^{\mu}\left(  \tau,\sigma\right)
d\sigma
\end{equation}%
\begin{equation}
X_{CM}^{\mu}=\frac{1}{\pi}\int_{0}^{\pi}x_{o}^{\mu}+\frac{1}{\pi T}p^{\mu}%
\tau+\frac{i}{\sqrt{\pi T}}\underset{n\neq0}{\sum}\frac{1}{n}\alpha_{n}^{\mu
}e^{-in\tau}\cos\left(  n\sigma\right)  d\sigma
\end{equation}%
\begin{equation}
X_{CM}^{\mu}=\frac{1}{\pi}\left[  x_{o}^{\mu}\pi+\frac{p^{\mu}}{\pi T}\tau
\pi+\frac{i}{\sqrt{\pi T}}\underset{n\neq0}{\sum}\frac{1}{n}\alpha_{n}^{\mu
}e^{-in\tau}\left.  \sin\left(  n\sigma\right)  \right|  _{o}^{\pi}\right]
\end{equation}%
\begin{equation}
X_{CM}^{\mu}=x_{o}^{\mu}+\frac{p^{\mu}}{\pi T}\tau
\end{equation}%

\begin{equation}
p_{CM}^{\mu}=\int_{0}^{\pi}\Pi^{\mu}d\sigma=T\int_{0}^{\pi}\overset{.}{x}%
^{\mu}d\sigma
\end{equation}%
\begin{equation}
p_{CM}^{\mu}=T\int_{0}^{\pi}\left[  \frac{1}{\pi T}p^{\mu}+\frac{1}{\sqrt{\pi
T}}\underset{n\neq0}{\sum}\alpha_{n}^{\mu}e^{-in\tau}\cos\left(
n\sigma\right)  \right]  d\sigma
\end{equation}%
\begin{equation}
p_{CM}^{\mu}=p^{\mu}%
\end{equation}

We can see that $x_{o}^{\mu}$ is the position of the center of mass \ at
$\tau=0$ and it moves as a free particle. The variables that describe the
motion of the string are $x_{o}^{\mu}$ and $p^{\mu}$ plus the collection of
$\alpha_{n}^{\mu}$. This means that the string moves as a whole (center of
mass) and vibrate in various modes.

\bigskip

\section{Classical Visaroro Algebra}

\bigskip

The equal-$\tau$ Poison Brackets for the $x^{\mu}$ and their conjugate momenta
$\Pi^{\mu}=T\overset{.}{x}^{\mu}$are:%

\begin{equation}
\left\{  x^{\mu}\left(  \sigma,\tau\right)  ,\overset{.}{x}^{\nu}\left(
\sigma^{\prime},\tau\right)  \right\}  _{PB}=\frac{1}{T}\eta^{\mu\nu}%
\delta\left(  \sigma-\sigma^{\prime}\right) \label{poison1}%
\end{equation}%

\begin{equation}
\left\{  x^{\mu}\left(  \sigma,\tau\right)  ,x^{\nu}\left(  \sigma^{\prime
},\tau\right)  \right\}  _{PB}=0
\end{equation}%

\begin{equation}
\left\{  \overset{.}{x}^{\mu}\left(  \sigma,\tau\right)  ,\overset{.}{x}^{\nu
}\left(  \sigma^{\prime},\tau\right)  \right\}  _{PB}=0\label{poison3}%
\end{equation}

\bigskip

Using the expresion for $x^{\mu}\left(  \sigma,\tau\right)  $ we get:%

\begin{equation}
\left\{  \alpha_{m}^{\mu},\alpha_{n}^{\nu}\right\}  =\left\{  \widetilde
{\alpha}_{m}^{\mu},\widetilde{\alpha}_{n}^{\nu}\right\}  =-im\eta^{\mu\nu
}\delta_{m+n,0}\label{poison bracket alfa1}%
\end{equation}%

\begin{equation}
\left\{  \widetilde{\alpha}_{m}^{\mu},\alpha_{n}^{\nu}\right\}
=0\label{poison bracket alfa2}%
\end{equation}%

\begin{equation}
\left\{  x_{0}^{\mu},p^{\nu}\right\}  =\eta^{\mu\nu}%
\label{poison bracket alfa3}%
\end{equation}

The Hamiltonian is give by:%

\begin{equation}
H=\int\left(  \overset{.}{x}^{\mu}\Pi_{\mu}-L\right)  d\sigma=T\int\left(
\overset{.}{x}^{\mu}\overset{.}{x}_{\mu}-L\right)  d\sigma
\end{equation}%

\begin{equation}
H=\int\left(  T\overset{.}{x}^{\mu}\overset{.}{x}_{\mu}+\frac{T}{2}%
\eta^{\alpha\beta}\eta_{\mu\nu}\partial_{\alpha}x^{\mu}\partial_{\beta}x^{\nu
}\right)  d\sigma
\end{equation}%

\begin{equation}
H=\int\left[  T\overset{.}{x}^{\mu}\overset{.}{x}_{\mu}+\frac{T}{2}\left(
-\overset{.}{x}^{\mu}\overset{.}{x}_{\mu}+x^{\prime\mu}x_{\mu}^{\prime
}\right)  \right]  d\sigma
\end{equation}%

\begin{equation}
H=\frac{T}{2}\int\left[  \overset{.}{x}^{\mu}\overset{.}{x}_{\mu}+x^{\prime
\mu}x_{\mu}^{\prime}\right]  d\sigma
\end{equation}

\bigskip

This hamiltonian can be expresed in terms of the oscillators in the closed
string case, using $\alpha_{0}^{\mu}=$\ $\frac{1}{\sqrt{\pi T}}p^{\mu}$ ,as:%

\begin{align*}
H  & =\frac{T}{2}\int\frac{1}{4\pi T}\left(  \underset{m}{\sum}\left(  \left(
\alpha_{m}^{\mu}\right)  ^{\dagger}e^{-im\sigma}+\left(  \widetilde{\alpha
}_{m}^{\mu}\right)  ^{\dagger}e^{im\sigma}\right)  e^{im\tau}\right)  \left(
\underset{n}{\sum}\left(  \left(  \alpha_{n}\right)  _{\mu}e^{in\sigma
}+\left(  \widetilde{\alpha}_{n}\right)  _{\mu}e^{-in\sigma}\right)
e^{-in\tau}\right) \\
& +\frac{1}{4\pi T}\left(  \underset{m\neq0}{\sum}\left(  -\left(  \alpha
_{m}^{\mu}\right)  ^{\dagger}e^{-im\sigma}+\left(  \widetilde{\alpha}_{m}%
^{\mu}\right)  ^{\dagger}e^{im\sigma}\right)  e^{im\tau}\right)  \left(
\underset{n\neq0}{\sum}\left(  -\left(  \alpha_{n}\right)  _{\mu}e^{in\sigma
}+\left(  \widetilde{\alpha}_{n}\right)  _{\mu}e^{-in\sigma}\right)
e^{-in\tau}\right)  d\sigma
\end{align*}%

\begin{align*}
H  & =\frac{T}{2}\int\frac{1}{4\pi T}\left(  \underset{m}{\sum}\left(
\alpha_{-m}^{\mu}e^{-im\sigma}+\widetilde{\alpha}_{-m}^{\mu}e^{im\sigma
}\right)  e^{im\tau}\right)  \left(  \underset{n}{\sum}\left(  \left(
\alpha_{n}\right)  _{\mu}e^{in\sigma}+\left(  \widetilde{\alpha}_{n}\right)
_{\mu}e^{-in\sigma}\right)  e^{-in\tau}\right) \\
& +\frac{1}{4\pi T}\left(  \underset{m\neq0}{\sum}\left(  -\alpha_{-m}^{\mu
}e^{-im\sigma}+\widetilde{\alpha}_{-m}^{\mu}e^{im\sigma}\right)  e^{im\tau
}\right)  \left(  \underset{n\neq0}{\sum}\left(  -\left(  \alpha_{n}\right)
_{\mu}e^{in\sigma}+\left(  \widetilde{\alpha}_{n}\right)  _{\mu}e^{-in\sigma
}\right)  e^{-in\tau}\right)  d\sigma
\end{align*}

Since:%

\begin{equation}
\int e^{-im\sigma}e^{in\sigma}d\sigma=2\pi\delta\left(  m-n\right)
\end{equation}%

\begin{equation}
\int e^{im\sigma}e^{in\sigma}d\sigma=2\pi\delta\left(  m+n\right)
\end{equation}

we obtain:

\bigskip%

\begin{align}
H  & =\frac{T}{2}\frac{1}{4\pi T}\left[  2\pi\underset{n}{\sum}\left(
\alpha_{-n}\cdot\alpha_{n}+\widetilde{\alpha}_{n}\cdot\alpha_{n}e^{-2in\tau
}+\alpha_{n}\cdot\widetilde{\alpha}_{n}e^{-2in\tau}+\widetilde{\alpha}%
_{-n}\cdot\widetilde{\alpha}_{n}\right)  \right. \nonumber\\
& \left.  +2\pi\underset{n\neq0}{\sum}\left(  \alpha_{-n}\cdot\alpha
_{n}-\widetilde{\alpha}_{n}\cdot\alpha_{n}e^{-2in\tau}-\alpha_{n}%
\cdot\widetilde{\alpha}_{n}e^{-2in\tau}+\widetilde{\alpha}_{-n}\cdot
\widetilde{\alpha}_{n}\right)  \right]
\end{align}%

\begin{equation}
H=\frac{T}{2}\frac{2\pi}{4\pi T}\left[  \alpha_{0}\cdot\widetilde{\alpha}%
_{0}+\widetilde{\alpha}_{0}\cdot\alpha_{0}+\underset{n}{\sum}\left(
\alpha_{-n}\cdot\alpha_{n}+\widetilde{\alpha}_{-n}\cdot\widetilde{\alpha}%
_{n}\right)  +\underset{n\neq0}{\sum}\left(  \alpha_{-n}\cdot\alpha
_{n}+\widetilde{\alpha}_{-n}\cdot\widetilde{\alpha}_{n}\right)  \right]
\end{equation}%

\begin{equation}
H=\frac{1}{4}\left[  2\left(  \alpha_{0}\cdot\alpha_{0}+\widetilde{\alpha}%
_{0}\cdot\widetilde{\alpha}_{0}\right)  +\underset{n\neq0}{\sum}\left(
\alpha_{-n}\cdot\alpha_{n}+\widetilde{\alpha}_{-n}\cdot\widetilde{\alpha}%
_{n}\right)  +\underset{n\neq0}{\sum}\left(  \alpha_{-n}\cdot\alpha
_{n}+\widetilde{\alpha}_{-n}\cdot\widetilde{\alpha}_{n}\right)  \right]
\end{equation}%

\begin{equation}
H=\frac{1}{4}\left[  2\left(  \alpha_{0}\cdot\alpha_{0}+\widetilde{\alpha}%
_{0}\cdot\widetilde{\alpha}_{0}\right)  +\underset{n\neq0}{\sum}\left(
\alpha_{-n}\cdot\alpha_{n}+\widetilde{\alpha}_{-n}\cdot\widetilde{\alpha}%
_{n}\right)  +\underset{n\neq0}{\sum}\left(  \alpha_{-n}\cdot\alpha
_{n}+\widetilde{\alpha}_{-n}\cdot\widetilde{\alpha}_{n}\right)  \right]
\end{equation}%

\begin{equation}
H=\frac{1}{4}\left[  \ 2\left(  \alpha_{0}\cdot\alpha_{0}+\widetilde{\alpha
}_{0}\cdot\widetilde{\alpha}_{0}\right)  +2\underset{n\neq0}{\sum}\left(
\alpha_{-n}\cdot\alpha_{n}+\widetilde{\alpha}_{-n}\cdot\widetilde{\alpha}%
_{n}\right)  \right]
\end{equation}%

\begin{equation}
H=\frac{1}{2}\left[  \ \left(  \alpha_{0}\cdot\alpha_{0}+\widetilde{\alpha
}_{0}\cdot\widetilde{\alpha}_{0}\right)  +\underset{n\neq0}{\sum}\left(
\alpha_{-n}\cdot\alpha_{n}+\widetilde{\alpha}_{-n}\cdot\widetilde{\alpha}%
_{n}\right)  \right]
\end{equation}%

\begin{equation}
H=\frac{1}{2}\underset{n}{\sum}\left(  \alpha_{-n}\cdot\alpha_{n}%
+\widetilde{\alpha}_{-n}\cdot\widetilde{\alpha}_{n}\right)
\label{open string hamiltonian}%
\end{equation}

\bigskip

\bigskip For the open string using again $\alpha_{o}^{\mu}=\frac{p^{\mu}%
}{\sqrt{\pi T}},$ we have:

\bigskip%
\begin{align*}
H  & =\frac{T}{2}\int\frac{1}{4\pi T}\left(  \underset{m}{\sum}\left(
\alpha_{m}^{\mu}\right)  ^{\dagger}\left(  e^{-im\sigma}+e^{im\sigma}\right)
e^{im\tau}\right)  \left(  \underset{n}{\sum}\left(  \alpha_{n}\right)  _{\mu
}\left(  e^{in\sigma}+e^{-in\sigma}\right)  e^{-in\tau}\right) \\
& +\frac{1}{4\pi T}\left(  \underset{m\neq0}{\sum}\left(  \alpha_{m}^{\mu
}\right)  ^{\dagger}\left(  -e^{-im\sigma}+e^{im\sigma}\right)  e^{im\tau
}\right)  \left(  \underset{n\neq0}{\sum}\left(  \alpha_{n}\right)  _{\mu
}\left(  -e^{in\sigma}+e^{-in\sigma}\right)  e^{-in\tau}\right)  d\sigma
\end{align*}%

\begin{align*}
H  & =\frac{T}{2}\int\frac{1}{4\pi T}\left(  \underset{m}{\sum}\alpha
_{-m}^{\mu}\left(  e^{-im\sigma}+e^{im\sigma}\right)  e^{im\tau}\right)
\left(  \underset{n}{\sum}\left(  \alpha_{n}\right)  _{\mu}\left(
e^{in\sigma}+e^{-in\sigma}\right)  e^{-in\tau}\right) \\
& +\frac{1}{4\pi T}\left(  \underset{m\neq0}{\sum}\alpha_{-m}^{\mu}\left(
-e^{-im\sigma}+e^{im\sigma}\right)  e^{im\tau}\right)  \left(  \underset
{n\neq0}{\sum}\left(  \alpha_{n}\right)  _{\mu}\left(  -e^{in\sigma
}+e^{-in\sigma}\right)  e^{-in\tau}\right)  d\sigma
\end{align*}%

\begin{align}
H  & =\frac{T}{2}\frac{1}{4\pi T}\left[  \pi\underset{n}{\sum}\left(
\alpha_{-n}\cdot\alpha_{n}+\alpha_{n}\cdot\alpha_{n}e^{-2in\tau}+\alpha
_{n}\cdot\alpha_{n}e^{-2in\tau}+\alpha_{-n}\cdot\alpha_{n}\right)  \right.
\nonumber\\
& \left.  +\pi\underset{n\neq0}{\sum}\left(  \alpha_{-n}\cdot\alpha_{n}%
-\alpha_{n}\cdot\alpha_{n}e^{-2in\tau}-\alpha_{n}\cdot\alpha_{n}e^{-2in\tau
}+\alpha_{-n}\cdot\alpha_{n}\right)  \right]
\end{align}%

\begin{equation}
H=\frac{T}{2}\frac{\pi}{4\pi T}\left[  \alpha_{0}\cdot\alpha_{0}+\alpha
_{0}\cdot\alpha_{0}+\underset{n}{\sum}\left(  \alpha_{-n}\cdot\alpha
_{n}+\alpha_{-n}\cdot\alpha_{n}\right)  +\underset{n\neq0}{\sum}\left(
\alpha_{-n}\cdot\alpha_{n}+\alpha_{-n}\cdot\alpha_{n}\right)  \right]
\end{equation}%

\begin{equation}
H=\frac{1}{8}\left[  2\left(  \alpha_{0}\cdot\alpha_{0}+\alpha_{0}\cdot
\alpha_{0}\right)  +\underset{n\neq0}{\sum}2\alpha_{-n}\cdot\alpha
_{n}+\underset{n\neq0}{\sum}2\alpha_{-n}\cdot\alpha_{n}\right]
\end{equation}%

\begin{equation}
H=\frac{1}{8}\left[  4\left(  \alpha_{0}\cdot\alpha_{0}\right)  +4\underset
{n\neq0}{\sum}\alpha_{-n}\cdot\alpha_{n}\right]
\end{equation}%

\begin{equation}
H=\frac{1}{2}\left[  \left(  \alpha_{0}\cdot\alpha_{0}\right)  +\underset
{n\neq0}{\sum}\alpha_{-n}\cdot\alpha_{n}\right]
\end{equation}%

\begin{equation}
H=\frac{1}{2}\underset{n}{\sum}\alpha_{-n}\cdot\alpha_{n}%
\label{closed string hamiltonian}%
\end{equation}

\bigskip

\bigskip TheVisaroro operators are defined \ as the Fourier modes of the
Stress tensor. For the closed string they are:%

\begin{equation}
L_{m}=\frac{1}{2}\int_{0}^{2\pi}T_{--}e^{im\left(  \tau-\sigma\right)
}d\sigma\text{ \ \ \ \ \ \ , \ \ \ \ \ \ }\widetilde{L}_{m}=\frac{1}{2}%
\int_{0}^{2\pi}T_{++}e^{im\left(  \tau+\sigma\right)  }d\sigma
\end{equation}

\bigskip Expressed in oscillators these operators are:%

\begin{equation}
L_{m}=\frac{1}{2}\underset{n}{\sum}\alpha_{m-n}\cdot\alpha_{n}\text{
\ \ \ \ \ \ , \ \ \ \ \ \ }\widetilde{L}_{m}=\frac{1}{2}\underset{n}{\sum
}\widetilde{\alpha}_{m-n}\cdot\widetilde{\alpha}_{n}%
\end{equation}

\bigskip

Because of the properties of the $\alpha$'s, the operators satisfy:%

\begin{equation}
L_{m}^{\dagger}=\frac{1}{2}\underset{n}{\sum}\alpha_{m-n}^{\dagger}\cdot
\alpha_{n}^{\dagger}=\frac{1}{2}\underset{n}{\sum}\alpha_{-m+n}\cdot
\alpha_{-n}=L_{-m}%
\end{equation}%

\begin{equation}
\widetilde{L}_{m}^{\dagger}=\frac{1}{2}\underset{n}{\sum}\widetilde{\alpha
}_{m-n}^{\dagger}\cdot\widetilde{\alpha}_{n}^{\dagger}=\frac{1}{2}\underset
{n}{\sum}\widetilde{\alpha}_{-m+n}\cdot\widetilde{\alpha}_{-n}=\widetilde
{L}_{-m}%
\end{equation}

In terms of the Visaroro operators, the Hamiltonian (\ref{open string
hamiltonian}) can be expresed as:%

\begin{equation}
H=L_{0}+\widetilde{L}_{0}%
\end{equation}

\bigskip

For the open string the $\widetilde{\alpha}$'s are equal to the $\alpha$'s,
then the Visaroro operators are definied as:%

\begin{equation}
L_{m}=\frac{1}{2}\int_{0}^{\pi}\left[  T_{--}e^{im\left(  \tau-\sigma\right)
}+T_{++}e^{im\left(  \tau+\sigma\right)  }\right]  d\sigma
\end{equation}

\bigskip or in terms of the oscillators:%

\begin{equation}
L_{m}=\frac{1}{2}\underset{n}{\sum}\alpha_{m-n}\cdot\alpha_{n}%
\end{equation}

\bigskip

In this case, the Hamiltonian (\ref{closed string hamiltonian}) can be written as:%

\begin{equation}
H=L_{0}%
\end{equation}

Using (\ref{poison bracket alfa1}),(\ref{poison bracket alfa2}) and
(\ref{poison bracket alfa3}) is easy to derive the Poison brackets for the
Visaroro operators:%

\begin{equation}
\left\{  L_{m},L_{n}\right\}  _{PB}=-i\left(  m-n\right)  L_{m+n}%
\end{equation}%

\begin{equation}
\left\{  \widetilde{L}_{m},\widetilde{L}_{n}\right\}  _{PB}=-i\left(
m-n\right)  \widetilde{L}_{m+n}%
\end{equation}%

\begin{equation}
\left\{  L_{m},\widetilde{L}_{n}\right\}  _{PB}=0
\end{equation}

For the open string the $\widetilde{L}$'s operators are absent.

\chapter{\bigskip Bosonic String Quantization}

\bigskip There exist three aproaches to quantize classical strings:

\begin{enumerate}
\item \textbf{Canonical Quantization:}

In this aproach the classical variables become operators. Because of the
constraints there are two options:

\begin{enumerate}
\item  Covariant Quantization

This procedure quantize first and then impose the constraints as conditions on
states in Hilbert space. Is named covariant because it preserves the Lorentz invariace.

\item  Light-Cone Quantization

Here, the \ constraints are solved in the classical system, so it leaves a
smaller number of classical variables. Then they are quantized. In this
aproach the manifest Lorentz invariance is lost.
\end{enumerate}

\item \textbf{Path Integral Quantization}

This way preserves the manifest Lorentz invarinace, but contains ghost fields.
\end{enumerate}

\bigskip

\section{Covariant Canonical Quantization}

In the canonical quantization the fields are replaced by operators, and the
Poisson brackets are reemplazed by commutators following the rule:%

\begin{equation}
\left\{  \text{ },\text{ }\right\}  _{PB}\longrightarrow-i\left[  \text{
},\text{ }\right]
\end{equation}

Following this rule, the relations (\ref{poison1})-(\ref{poison3}) become:%

\begin{equation}
\left[  x^{\mu}\left(  \sigma,\tau\right)  ,\overset{.}{\Pi}^{\nu}\left(
\sigma^{\prime},\tau\right)  \right]  =i\eta^{\mu\nu}\delta\left(
\sigma-\sigma^{\prime}\right)
\end{equation}%

\begin{equation}
\left[  x^{\mu}\left(  \sigma,\tau\right)  ,x^{\nu}\left(  \sigma^{\prime
},\tau\right)  \right]  =0
\end{equation}%

\begin{equation}
\left[  \overset{.}{\Pi}^{\mu}\left(  \sigma,\tau\right)  ,\overset{.}{\Pi
}^{\nu}\left(  \sigma^{\prime},\tau\right)  \right]  =0
\end{equation}

\bigskip Or in terms of the oscillator we can write:%

\begin{equation}
\left[  \alpha_{m}^{\mu},\alpha_{n}^{\nu}\right]  =\left[  \widetilde{\alpha
}_{m}^{\mu},\widetilde{\alpha}_{n}^{\nu}\right]  =m\eta^{\mu\nu}\delta_{m+n,0}%
\end{equation}%

\begin{equation}
\left[  \widetilde{\alpha}_{m}^{\mu},\alpha_{n}^{\nu}\right]  =0
\end{equation}%

\begin{equation}
\left[  x_{0}^{\mu},p^{\nu}\right]  =i\eta^{\mu\nu}%
\end{equation}

The first of this conditions can be written, absorbing the factor $m$ as:%

\begin{equation}
\left[  \alpha_{m}^{\mu},\left(  \alpha_{n}^{\nu}\right)  ^{\dagger}\right]
=\left[  \widetilde{\alpha}_{m}^{\mu},\left(  \widetilde{\alpha}_{n}^{\nu
}\right)  ^{\dagger}\right]  =\eta^{\mu\nu}\delta_{m,n}%
\end{equation}

In Quantum Mechanics the harmonic oscillator can be described using raising
and lowering operatos that satisfy the condition:%

\begin{equation}
\left[  \alpha,\alpha^{\dagger}\right]  =1
\end{equation}

\bigskip

Then, we see that the expansion coefficients $\alpha_{-n}^{\mu}=$ $\left(
\alpha_{n}^{\mu}\right)  ^{\dagger}$ and $\alpha_{n}^{\mu}$ are raising and
lowering operators respectively. But there are one problem. Because of the
metric tensor sign: $\eta^{00}=-1$ the time component of the oscillators has a
minus sign. this means that they create states of negative norm, and it makes
the quantum theory inconsistent. However, as we will see the classical
condition $T_{\alpha\beta}=0$ will eliminate the negative-norm states from
physical spectrum.

\bigskip

Since the system is now an infinite set of harmonic oscillators, we may define
the Hilbert space in a simple way. The ground-state is the one that is
anihilated by all the lowering operators, but since this state is not
completly determined by this, we use the center-of-mass momentum operator.
Diagonalizing $p^{\mu}$ the states are caracterized by its momentum. So, the
ground state may be written as:%

\begin{equation}
\alpha_{m}\left|  0,p\right\rangle =0\text{ \ \ \ \ \ \ }\forall m>0
\end{equation}

\bigskip

Other states are obtained by aplying the raising operators:%

\begin{equation}
\alpha_{-1}^{\mu}\left|  0,p\right\rangle \text{ , \ }\alpha_{-2}^{\mu}\left|
p\right\rangle \text{, \ }\underset{m}{\prod}\alpha_{m}^{\mu}\left|
0,p\right\rangle
\end{equation}

\bigskip

\subsection{Visaroro Operators}

In order to define the Visaroro operators in the quantum system we have to
introduce the normal ordering operation that puts all positive frequency modes
(lowering operators) to the right of the negative frequency modes (raising operators).

With this definition the Visaroro operators are defined as :%

\begin{equation}
L_{m}=\frac{1}{2}\underset{n}{\sum}:\alpha_{m-n}\cdot\alpha_{n}:
\end{equation}

and similarly for the closed string:%

\begin{equation}
\widetilde{L}_{m}=\frac{1}{2}\underset{n}{\sum}:\widetilde{\alpha}_{m-n}%
\cdot\widetilde{\alpha}_{n}:
\end{equation}

From now on, for the closed string there are similar relations for the
operators $\widetilde{L}_{m}$.

\bigskip

The only operator affected by the normal ordering is $L_{0}$:%

\begin{equation}
L_{0}=\frac{1}{2}\alpha_{0}\cdot\alpha_{0}+\frac{1}{2}\underset{n}{\sum
}:\alpha_{-n}\cdot\alpha_{n}:
\end{equation}%

\begin{equation}
L_{0}=\frac{1}{2}\alpha_{0}\cdot\alpha_{0}+\underset{n=1}{\overset{\infty
}{\sum}}\alpha_{-n}\cdot\alpha_{n}\label{lo operator}%
\end{equation}

Now, because of the arbitrariness in definig the normal ordering we have to
include a normal-ordering constant $q$ in the expresion for $L_{0}$.%

\begin{equation}
L_{0}\longrightarrow L_{0}-q
\end{equation}

\bigskip

The visaroro algebra can be calculated using the definition of the operators,
and it gives ( for a detailed calculation see \cite{Scherk}):

\bigskip%

\begin{equation}
\left[  L_{m},L_{n}\right]  =\left(  m-n\right)  L_{m+n}+\frac{c}{12}m\left(
m^{2}-1\right)  \delta_{m+n,0}\label{visaroro algebra}%
\end{equation}

\bigskip

Its important to note that there exist a difference between the operator and
the Poisson algebra. this diference is a c-number which depends on the
dimension of space-time, since in the case we have $c=d$. This term is called
the ''conformal anomaly term'' and the cnstant $c$ is in general called the
''central charge''. This term is an inescapable consequence of the
quantization, and is because of this that the string theory depends so much on
the dimension of space-time.

\bigskip

The classical Visaroro constraints (\ref{visaroro constraint}) cannot be
imposed as operators constraints $L_{m}\left|  \psi\right\rangle =0$, because
(\ref{visaroro algebra}) gives:%

\begin{equation}
\left\langle \psi\right|  \left[  L_{m},L_{-m}\right]  \left|  \psi
\right\rangle =2m\left\langle \psi\right|  L_{0}\left|  \psi\right\rangle
+\frac{d}{12}m\left(  m^{2}-1\right)  \left\langle \psi\right|  \left.
\psi\right\rangle \neq0
\end{equation}

\bigskip

Thus, the constraints are imposed as:%

\begin{equation}
\left(  L_{0}-q\right)  \left|  \phi\right\rangle
=0\label{quantum cosntraint 1}%
\end{equation}%

\begin{equation}
L_{m}\left|  \phi\right\rangle =0\text{ \ \ \ \ \ \ \ \ \ }m>0
\end{equation}

where $\left|  \phi\right\rangle $ denotes a physical state. This is
consistent with the classical constraints because:%

\begin{equation}
\left\langle \phi^{\prime}\right|  \left(  L_{m}-q\delta_{m,0}\right)  \left|
\phi\right\rangle =0\text{ \ \ \ \ \ \ \ \ \ \ \ \ \ \ \ \ \ \ \ \ for all }m
\end{equation}

As we will see, the constraint for $L_{0}$ is a generalization of the
Klein-Gordon equation since it contains a term with $p^{2}=-\partial
\cdot\partial$ and a term that determines the mass of the state.

\bigskip

\subsection{Bosonic String Spectrum}

\bigskip\textbf{Open String.}

Using the form of $L_{0}$ given by (\ref{lo operator}), the constraint
(\ref{quantum cosntraint 1}) gives:%

\begin{equation}
\frac{1}{2}\alpha_{0}\cdot\alpha_{0}+\underset{n}{\sum}\alpha_{-n}\cdot
\alpha_{n}=q
\end{equation}%

\begin{equation}
\frac{1}{2}\alpha_{0}\cdot\alpha_{0}=-\underset{n}{\sum}\alpha_{-n}\cdot
\alpha_{n}+q
\end{equation}%

\begin{equation}
\frac{1}{2\pi T}p^{\mu}p_{\mu}=-\underset{n}{\sum}\alpha_{-n}\cdot\alpha_{n}+q
\end{equation}

\bigskip And using the mass-shell condition $p^{\mu}p_{\mu}=-m^{2}$ , we have:%

\begin{equation}
\frac{1}{2\pi T}m^{2}=N-q\label{mass shell condition}%
\end{equation}

\bigskip

where is defined the level-number operator as:%

\begin{equation}
N=\underset{n}{\sum}\alpha_{-n}\cdot\alpha_{n}%
\end{equation}

Now, we will set $q=1$ since this value is necessary for a consistent theory
(as we will se later). Since each $\alpha_{-n}\cdot\alpha_{n}$ has eiegnvalues
$0,1,2,3,...$ etc. then the number operator \ also has values $0,1,2,...$.
\ The ground state, $\left|  0,p\right\rangle $ is given by $N=0$ . So we have:%

\begin{equation}
\frac{1}{2\pi T}m^{2}=-1
\end{equation}

This state corresponds to a Tachyon (since $p^{\mu}$ is space-like). This
particle may cause a vacuum unstability, then it is not possible in a
consisten quantum theory. Anyway, we will not worry about it now and will
continue with the spectum analisis.

\bigskip

The first excited state has $N=1$ and (\ref{mass shell condition}) gives $m=0
$. To obtain this sate we need:%

\begin{equation}
\left|  \phi\right\rangle =\xi_{\mu}\alpha_{-1}^{\mu}\left|  0,p\right\rangle
\end{equation}

\bigskip

where $\xi^{\mu}$ is the polarization vector for a massless spin-1 particle.
Now, the Visaroro constraint gives:%

\begin{equation}
L_{1}\left|  \phi\right\rangle =0
\end{equation}%

\begin{equation}
\xi_{\mu}L_{1}\alpha_{-1}^{\mu}\left|  0,p\right\rangle =0
\end{equation}

Since the operators $\alpha_{m}$ $\left(  m>0\right)  $ anihilates the
ground-state, $\ L_{1}$ has just the contributions:%

\begin{equation}
\xi_{\mu}\left(  \frac{1}{2}\underset{n}{\sum}:\alpha_{1-n}\cdot\alpha
_{n}:\right)  \alpha_{-1}^{\mu}\left|  0,p\right\rangle =0
\end{equation}%

\begin{equation}
\frac{1}{2}\xi_{\mu}\left(  \alpha_{0}\cdot\alpha_{1}+\alpha_{1}\cdot
\alpha_{0}\right)  \alpha_{-1}^{\mu}\left|  0,p\right\rangle =0
\end{equation}%

\begin{equation}
\xi_{\mu}\left(  \alpha_{0}\right)  _{\nu}\alpha_{1}^{\nu}\alpha_{-1}^{\mu
}\left|  0,p\right\rangle =0
\end{equation}%

\begin{equation}
\xi_{\mu}\left(  \alpha_{0}\right)  _{\nu}\left[  \alpha_{-1}^{\mu}\alpha
_{1}^{\nu}+\eta^{\mu\nu}\right]  \left|  0,p\right\rangle =0
\end{equation}%

\begin{equation}
\xi_{\mu}\left(  \alpha_{0}\right)  _{\nu}\alpha_{-1}^{\mu}\alpha_{1}^{\nu
}\left|  0,p\right\rangle +\xi_{\mu}\left(  \alpha_{0}\right)  _{\nu}\eta
^{\mu\nu}\left|  0,p\right\rangle =0
\end{equation}%

\begin{equation}
\xi_{\mu}\left(  \alpha_{0}\right)  ^{\mu}\left|  0,p\right\rangle =0
\end{equation}%

\begin{equation}
\xi_{\mu}\frac{1}{\sqrt{\pi T}}p^{\mu}\left|  0,p\right\rangle =0
\end{equation}%

\begin{equation}
\xi_{\mu}p^{\mu}\left|  0,p\right\rangle =0
\end{equation}

\bigskip Then , the Visaroro constraint implies that $\xi_{\mu}$ must satisfy:%

\begin{equation}
p^{\mu}\xi_{\mu}=0
\end{equation}

This condition ensures that the spin is transversely polarized, so there are
just $d-2$ independient polarization states.

The second excited state, with $N=2$ , has:%

\begin{equation}
\frac{1}{2\pi T}m^{2}=1
\end{equation}

\bigskip The most general state is a superposition :%

\begin{equation}
\left|  \phi\right\rangle =\left(  \xi_{\mu}\alpha_{-2}^{\mu}+\lambda_{\mu\nu
}\alpha_{-1}^{\mu}\alpha_{-1}^{\nu}\right)  \left|  0,p\right\rangle
\end{equation}

And again, the Visaroro constraints%

\begin{equation}
L_{1}\left|  \phi\right\rangle =L_{2}\left|  \phi\right\rangle =0
\end{equation}

restrict the values of $\xi_{\mu}$ and $\lambda_{\mu\nu}$.

\bigskip

\textbf{Closed String.}

\bigskip

For the closed string we have the conditions:%

\begin{equation}
\left(  L_{0}-q\right)  \left|  \phi\right\rangle =0\label{closed 1}%
\end{equation}%

\begin{equation}
\left(  \widetilde{L}_{0}-q\right)  \left|  \phi\right\rangle
=0\label{closed 2}%
\end{equation}

the sum of this conditions gives:%

\begin{equation}
\left(  L_{0}+\widetilde{L}_{0}-2q\right)  \left|  \phi\right\rangle =0
\end{equation}

\bigskip So;
\begin{equation}
\frac{1}{2}\alpha_{0}\cdot\alpha_{0}+\underset{n}{\sum}\alpha_{-n}\cdot
\alpha_{n}+\frac{1}{2}\widetilde{\alpha}_{0}\cdot\widetilde{\alpha}%
_{0}+\underset{n}{\sum}\widetilde{\alpha}_{-n}\cdot\widetilde{\alpha}_{n}=2q
\end{equation}%

\begin{equation}
\alpha_{0}\cdot\alpha_{0}=-\underset{n}{\sum}\alpha_{-n}\cdot\alpha
_{n}-\underset{n}{\sum}\widetilde{\alpha}_{-n}\cdot\widetilde{\alpha}_{n}+2q
\end{equation}%

\begin{equation}
\frac{1}{\pi T}p^{\mu}p_{\mu}=-\underset{n}{\sum}\alpha_{-n}\cdot\alpha
_{n}-\underset{n}{\sum}\widetilde{\alpha}_{-n}\cdot\widetilde{\alpha}_{n}+2q
\end{equation}

\bigskip Using the mass-shell condition $p^{\mu}p_{\mu}=-m^{2}$ , we have:%

\begin{equation}
\frac{1}{\pi T}m^{2}=N+\widetilde{N}-2q\label{mass shell condition closed}%
\end{equation}

\bigskip

where the level-number operators are:%

\begin{equation}
N=\underset{n}{\sum}\alpha_{-n}\cdot\alpha_{n}%
\end{equation}%

\begin{equation}
\widetilde{N}=\underset{n}{\sum}\widetilde{\alpha}_{-n}\cdot\widetilde{\alpha
}_{n}%
\end{equation}

\bigskip The difference between (\ref{closed 1}) and (\ref{closed 2}) gives:%

\begin{equation}
\left(  L_{0}-\widetilde{L}_{0}\right)  \left|  \phi\right\rangle =0
\end{equation}

Then we have the level-matching condition:%

\begin{equation}
L_{0}=\widetilde{L}_{0}%
\end{equation}%

\begin{equation}
\frac{1}{2}\alpha_{0}\cdot\alpha_{0}+\underset{n}{\sum}\alpha_{-n}\cdot
\alpha_{n}=\frac{1}{2}\widetilde{\alpha}_{0}\cdot\widetilde{\alpha}%
_{0}+\underset{n}{\sum}\widetilde{\alpha}_{-n}\cdot\widetilde{\alpha}_{n}%
\end{equation}%

\begin{equation}
\underset{n}{\sum}\alpha_{-n}\cdot\alpha_{n}=\underset{n}{\sum}\widetilde
{\alpha}_{-n}\cdot\widetilde{\alpha}_{n}%
\end{equation}%

\begin{equation}
N=\widetilde{N}%
\end{equation}

The closed string states are then the tensor product of left moving and right
moving states subject to the level matching condition. Then, the ground state
for the closed string is (without displaying the momuentum of the state):%

\begin{equation}
\left|  0\right\rangle \otimes\left|  0\right\rangle
\end{equation}

Since in the ground state has $N=\widetilde{N}=0,$ then, setting $q=1$ we obtain:%

\begin{equation}
\frac{1}{2\pi T}m^{2}=-2
\end{equation}

Again we obtain a spin 0 tachyon that makes the theory unststable.

The first excited state is:%

\begin{equation}
\left|  \phi\right\rangle =\xi_{\mu\nu}\left(  \alpha_{-1}^{\mu}\left|
0\right\rangle \otimes\widetilde{\alpha}_{-1}^{\nu}\left|  0\right\rangle
\right)
\end{equation}

\bigskip

This state has $m^{2}=0$.\ Again the Visaroro constraints%

\begin{equation}
L_{1}\left|  \phi\right\rangle =\widetilde{L}_{1}\left|  \phi\right\rangle =0
\end{equation}

restricts the possible values of the polarization to:%

\begin{equation}
p^{\mu}\xi_{\mu\nu}=0
\end{equation}

\bigskip

This kind of tensor gives three distinct spin states: the symmetric part
encodes a massless spin-2 particle $\left(  g_{\mu\nu}\right)  $, the
antisymmetric part is a massless antisymmetric tensor gauge field $B_{\mu\nu
}=-B_{\nu\mu}$ ; and finally there is encoded also an scalar field $\Phi.$

\bigskip

\bigskip

\bigskip

\bigskip

\backmatter

\chapter{Conclusion}

We have described some of the basic ideas of the 26-dimensional bosonic string
theory. There exist different kinds of bosonic strings, but all of them are
sick, because in each case the spectra contains a tachyon. This means that the
vacuum state is unstable, and it is still an open problem.The most remarkable
feature of string theories is their dimensionality. Here we have seen that 26
dimensions are needed and there is no consistent explication of why this
number is prefered.

Anyway the study of this toy model give a better understanding of the
different process and features that the final theory should or should not
have, and its a necesary step for the study of more elaborated models, sucha
as the Superstings.

\appendix

\bigskip

\bigskip
\end{document}